\PassOptionsToPackage{pdfpagelabels=false}{hyperref}
\documentclass[fleqn,usenatbib]{mnras}

\usepackage{mathptmx}
\usepackage{txfonts}

\usepackage[T1]{fontenc}
\usepackage{ae,aecompl}

\usepackage{graphicx}	

\newcommand{\mincir}{\raise -2.truept\hbox{\rlap{\hbox{$\sim$}}\raise5.truept
\hbox{$<$}\ }}
\newcommand{\magcir}{\raise -2.truept\hbox{\rlap{\hbox{$\sim$}}\raise5.truept
\hbox{$>$}\ }}
\newcommand{\siml}{\raise -2.truept\hbox{\rlap{\hbox{$\sim$}}\raise5.truept
\hbox{$<$}\ }}
\newcommand{\simg}{\raise -2.truept\hbox{\rlap{\hbox{$\sim$}}\raise5.truept
\hbox{$>$}\ }}
\newcommand{\be}{\begin{equation}}
\newcommand{\ee}{\end{equation}}
\newcommand{\ba}{\begin{eqnarray}}
\newcommand{\dagg} {\dag$\;$}
\newcommand{\ea}{\end{eqnarray}}
\newcommand {\kpc} {$h_{70}^{-1}$ kpc $\;$}

\newcommand {\h} {$h_{70}^{-1}$ Mpc$\;$}

\newcommand {\hhh} {\;h_{70}^{-1} \mathrm{Mpc}}
\newcommand {\ks} {km~s$^{-1} \;$}
\newcommand {\kss} {km~s$^{-1}$}
\newcommand {\m} {$M_{\odot} \;$}

\newcommand {\ml} {$h_{70}\;M_{\odot}/L_{\odot} \;$}

\newcommand{\degree}{\ensuremath{\mathrm{^\circ}}}

\newcommand{\arcs}{\ensuremath{\arcmm\hskip -0.1em\arcmm \;}}
\newcommand{\arcmm}{\ensuremath{\mathrm{^\prime}}}

\title{Multi-object spectroscopy of CL1821+643: a dynamically relaxed cluster with a giant radio halo?}

\author[W. Boschin et al.]{W. Boschin,$^{1,2,3}$\thanks{E-mail: boschin@tng.iac.es}
and M. Girardi$^{4,5}$
\\
$^{1}$Fundaci\'on G. Galilei - INAF (Telescopio Nazionale Galileo),
  Rambla J. A. Fern\'andez P\'erez 7, E-38712 Bre\~na Baja,
  Spain\\
$^{2}$Instituto de Astrof\'{\i}sica de Canarias, C/V\'{\i}a L\'actea
  s/n, E-38205 La Laguna, Spain\\
$^{3}$Departamento de Astrof\'{\i}sica, Univ. de La Laguna, Av. del
  Astrof\'{\i}sico Francisco S\'anchez s/n, E-38205 La Laguna, Spain\\
$^{4}$Dipartimento di Fisica dell'Universit\`a degli Studi di Trieste
  - Sezione di Astronomia, via Tiepolo 11, I-34143 Trieste, Italy\\
$^{5}$INAF - Osservatorio Astronomico di Trieste, via Tiepolo 11,
  I-34143 Trieste, Italy\\
}

\date{Accepted: 2018 July 9. Received: 2018 July 9; in original form: 2018 May 28}
\pubyear{2018}

\begin{document}
\label{firstpage}
\pagerange{\pageref{firstpage}--\pageref{lastpage}}
\maketitle

\begin{abstract}

We present the study of the dynamical status of the galaxy cluster
CL1821+643, a rare and intriguing cool-core cluster hosting a giant
radio halo. We base our analysis on new spectroscopic data for 129
galaxies acquired at the Italian Telescopio Nazionale {\it
Galileo}. We also use spectroscopic data available from the literature
and photometric data from the Sloan Digital Sky Survey. We select 120
cluster member galaxies and compute the cluster redshift
$\left<z\right>\sim 0.296$ and the global line-of-sight velocity
dispersion $\sigma_{\rm V}\sim 1100$ \kss. The results of our analysis
are consistent with CL1821+643 being a massive ($M>10^{15}$\m)
dynamically relaxed cluster dominated by a big and luminous elliptical
at the centre of the cluster potential well. None of the tests
employed to study the cluster galaxies kinematics in the 1D (velocity
information), 2D (spatial information), and 3D (combined velocity and
spatial information) domains is able to detect significant
substructures. While this picture is in agreement with previous
results based on X-ray data and on the existence of the central cool
core, we do not find any evidence of a merging process responsible for
the radio halo discovered in this cluster. Thus, this radio halo
remains an open problem that raises doubts about our understanding of
diffuse radio sources in clusters.

\end{abstract}


%

\section{INTRODUCTION}
\label{intro}

A fraction of the most massive galaxy clusters exhibits in their inner
regions diffuse radio sources named {\it radio haloes} (also {\it
giant radio haloes}, or GRHs). They are extended (on $~$1 Mpc scales)
unpolarized synchrotron sources produced by relativistic electrons and
large-scale magnetic fields spread out the intracluster medium (ICM),
not associated with compact radio sources like radio galaxies (see
e.g., Feretti et al. \citeyear{fer12} for a review).

The existence of GRHs indicates the presence in the ICM of particle
acceleration mechanisms. In particular, GRHs are probably related to
turbulence induced in the ICM by recent mergers (e.g. Brunetti \&
Jones \citeyear{bru15}). In fact, for several decades after their
discovery, these sources have always been found in dynamically
disturbed systems, i.e. massive merging clusters characterized by the
absence of a cool core (Feretti et al. \citeyear{fer12} and references
therein).

This picture has dramatically changed in 2014, when Bonafede et
al. (\citeyear{bon14}, hereafter B14) published the discovery of a GRH
in the cool-core cluster CL1821+643. More recently, Sommer et
al. (\citeyear{som17}) found diffuse radio emission at Mpc scale on
two $z\sim 0.22$ relaxed clusters, Abell 2390 and Abell 2261. Another
case is Abell 2142 ($z\sim 0.09$), where Venturi et
al. (\citeyear{ven17}) surprisingly found a GRH despite the fact that
its optical and X-ray properties do not reveal a major
merger. Finally, Savini et al. (\citeyear{sav18}) reported the
detection of diffuse radio emission on scales larger than the core in
the cool-core cluster PSZ1G139.61+24 ($z\sim 0.27$).

In summary, these recent findings suggest the possible existence of a
new class of dynamically relaxed clusters with GRHs. In the framework
of turbulent models this is surprising and challenges the idea that
major mergers, necessary to power GRHs, always disrupt the cool core.

In this paper, we focus on one of the puzzling clusters cited above:
CL1821+643 (hereafter CL1821). CL1821 hosts an extended central
diffuse emission with linear size $\sim 1.1\hhh$ and an extrapolated
radio power $P_{\rm 1.4\ GHz} \sim (3.6-3.8)\ 10^{24}$ W Hz$^{-1}$
elongated along the SE--NW direction (see B14). Because of its
location and size, B14 classified this source as a GRH.

\begin{figure*}
\centering 
\includegraphics[width=18cm]{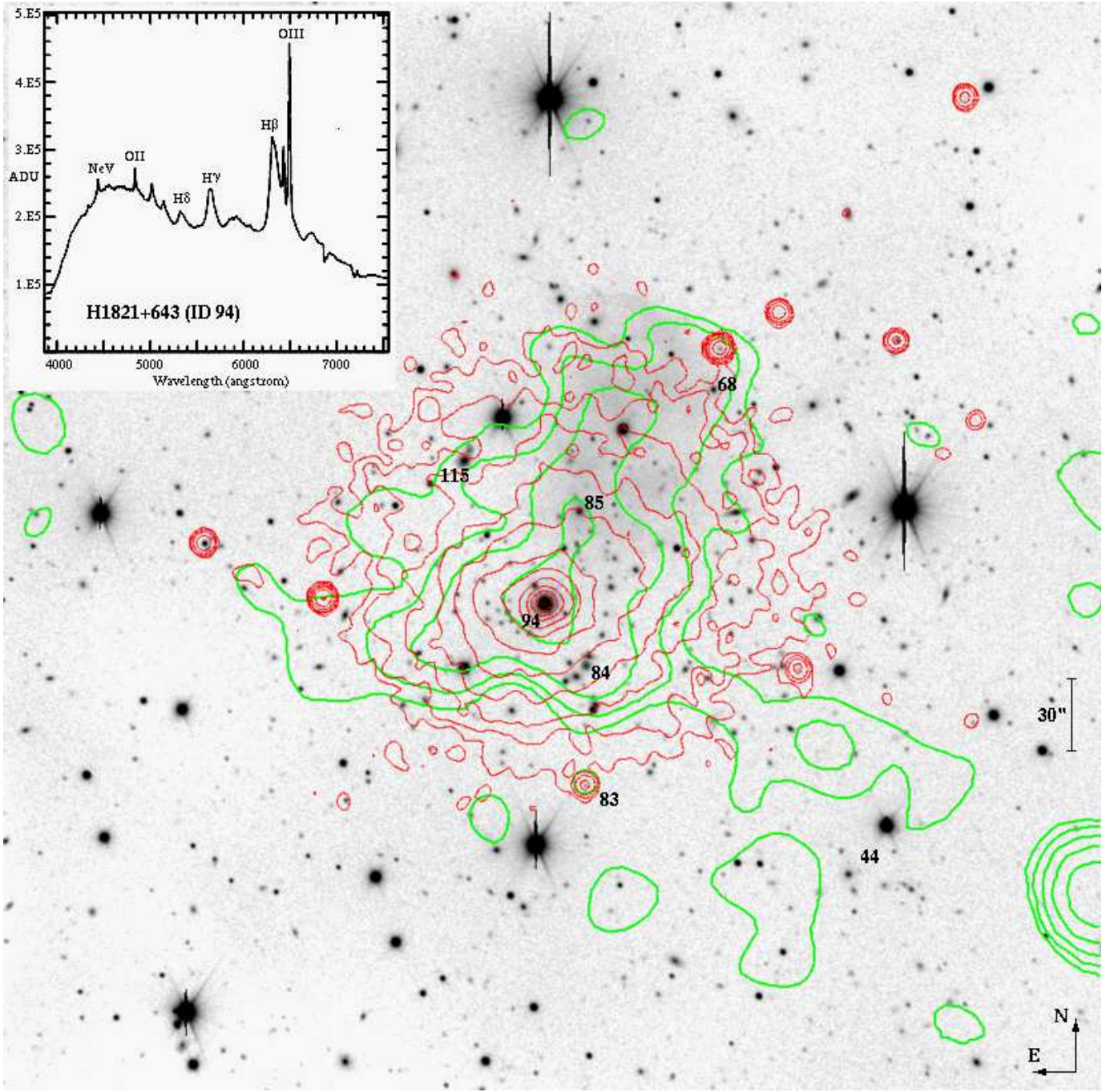}
\caption{The cluster CL1821 in the optical, X-ray and radio bands. The TNG grey--scale image (FOV $\sim$8 arcmin $\times$ 8 arcmin) in the background corresponds to the optical $r'$-band. Red thin contours show the cluster X-ray emission in the 0.5--7 keV band (from \textit{Chandra} archival image ID~9398; Texp: 34 ks). Thick green contours are the contour levels of a GMRT 323 MHz low-resolution image and show the radio halo (from B14). Top left hand inset is the TNG wavelength-calibrated spectrum of the central quasar H1821+643 (ID~94; see Table~\ref{catalogCL1821}). Other labels indicate galaxies mentioned in the text. The diffuse emission visible in the optical image NW of the galaxy ID~85 is the planetary nebula PN G094.0+27.4.}
\label{figimage}
\end{figure*}

From the optical point of view, CL1821 (at $z\sim$ 0.299, Schneider et
al. \citeyear{sch92}) is dominated by the central quasar H1821+643
(e.g. Hutchings \& Neff \citeyear{hut91}; Aravena et
al. \citeyear{ara11}; Reynolds et al. \citeyear{rey14}; Walker et
al. \citeyear{wal14}), an uncommon example of an optically very
luminous quasar hosted by the central dominant elliptical galaxy of a
massive cluster (see Fig.~\ref{figimage}). Moreover, CL1821 appears
slightly elongated in the SE--NW direction, as inferred from its weak
lensing properties (Wold et al. \citeyear{wol02}). However, our
present knowledge on this cluster is mainly based on X-ray
data. Russell et al. (\citeyear{rus10}) analysed {\it Chandra} data
taken with the ACIS-S instrument to study the ICM properties. In
particular, they found that the ICM temperature drops from $\sim$9 to
$\sim$1 keV, with a short central cooling time of $\sim$ 1 Gyr typical
of a relaxed strong cool-core cluster. They also concluded that the
quasar did not have a strong impact on the large-scale ICM properties.

The ACIS-S images suggest a projected morphology elongated on the
SE--NW direction (a feature revealed also by radio and weak lensing
data, see above and Fig.~\ref{figimage}) but do not show any evidence
of a major merger in this cluster similar to those detected in other
clusters with GRHs. Moreover, an analysis of CL1821 based on classical
X-ray morphological estimators (concentration parameter $c$ and power
ratio $P_3/P_0$; see B14 and references therein) also supports the
hypothesis that it is a relaxed galaxy system. Only the estimator $w$
(centroid shift parameter), could suggest the existence of an
undergoing minor or off-axis merger (see B14).

A more recent work by Kale and Parekh (\citeyear{kal16}) partially
disagrees with the findings of B14, claiming that a morphological
analysis of the same {\it Chandra} data (based on the parameters Gini,
$M_{20}$, and $c$; see Parekh et al. \citeyear{par15}) puts CL1821 in
the category of relaxed/non-relaxed clusters when H1821+643 is
included/excluded in/from the analysis.

Nevertheless, the X-ray morphological indicators cited above are not
sensitive to eventual mergers along the line of sight (LOS). The only
way to explore this possibility and to finally assess the dynamical
status of this intriguing cluster is to perform spectroscopic
observations of the cluster member galaxies. In fact, the spatial and
kinematical analysis of member galaxies constitute an effective tool
to detect substructures in clusters and highlight eventual pre-merging
subgroups or merger remnants (e.g. Boschin et al. \citeyear{bos04};
Boschin et al. \citeyear{bos13}). The optical information complements
the results of X-ray studies, also considering that the collisional
and non-collisional components of clusters (ICM and galaxies,
respectively) exhibit different behaviours during mergers (see e.g.
simulations by Roettiger et al. \citeyear{roe97}).

At the moment only a few dozen of galaxies have known redshifts in an
area 30 arcmin wide around H1821+643. Therefore, we decided to perform a
spectroscopic survey of CL1821, mainly sampling the central $\sim$1
Mpc size region of the cluster characterized by the diffuse X-ray and
radio emissions. In particular, we obtained new spectroscopic data at
the Italian Telescopio Nazionale {\em Galileo} (TNG), whose facilities
are well suited for the study of a galaxy cluster at $z\sim 0.3$ like
CL1821 (see e.g. the DARC project, Girardi et al. \citeyear{gir11}
and references therein \footnote{see also
http://wwwuser.oats.inaf.it/girardi/darc, the web site of the DARC
project.}).

This paper is organized as follows. We describe the optical
observations and present our spectroscopic data catalogue in
Sect.~\ref{data}. In Sect.~\ref{optanalysis} we explain the results of
our analysis of the cluster structure. We discuss our results and
present the conclusions in Sect.~\ref{disc}.

Throughout this paper, we use $H_0=70$ km s$^{-1}$ Mpc$^{-1}$ and
$h_{70}=H_0/(70$ km s$^{-1}$ Mpc$^{-1}$) in a flat cosmology with
$\Omega_0=0.3$ and $\Omega_{\Lambda}=0.7$. In the adopted cosmology, 1
arcmin corresponds to $\sim 265$ \kpc at the cluster redshift.  Unless
otherwise stated, we indicate errors at the 68\% confidence level
(hereafter c.l.).

\section{OBSERVATIONS AND BUILDING OF THE GALAXY CATALOGUE}
\label{data}

\subsection{TNG optical observations}
\label{optdata}

We used the instrument DOLORES of the TNG to perform multi-object
spectroscopic (MOS) observations of CL1821 in 2016 June (program
CAT16A\textunderscore 15). In particular, we observed four MOS masks
with the LR-B Grism\footnote{http://www.tng.iac.es/instruments/lrs}
obtaining spectra for 155 objects. The total exposure times were 5400s
for three masks and 3600s for the last one.

We reduced the spectra with standard {\sevensize
IRAF}\footnote{{\sevensize IRAF} is distributed by the National
Optical Astronomy Observatories, which are operated by the Association
of Universities for Research in Astronomy, Inc., under cooperative
agreement with the National Science Foundation.} tasks and were able
to compute redshifts for 123 galaxies by using the classic
cross-correlation technique (Tonry \& Davis \citeyear{ton79}). For
another six galaxies (IDs. 11, 31, 67, 72, 113, and 180, see
Table~\ref{catalogCL1821}) we estimated the redshifts by measuring the
wavelength location of prominent emission lines in their spectra (see
details of the redshift computation and their errors in Boschin et
al. \citeyear{bos13}).

\subsection{Spectroscopic catalogue and prominent galaxies}
\label{prom}

We also explored the NASA/IPAC Extragalactic Database
  (NED)\footnote{NASA/IPAC Extragalactic Database is operated by the
  Jet Propulsion Laboratory, California Institute of Technology, under
  contract with the National Aeronautics and Space Administration.}
  to search for galaxies with known redshift lying in the field
  covered by our spectroscopic survey. In total, we found a set of 72
  galaxies with measured redshift. 14 of these galaxies are in common
  with our TNG data. Since there is no evidence of systematic
  deviations between these 14 literature redshifts and our
  measurements, we added the remaining 58 NED galaxies to our sample.

Our final spectroscopic catalogue includes 187 galaxies in the field
of CL1821. For all the galaxies, we also use photometric information
from the Sloan Digital Sky Survey (SDSS) DR13 in the magnitude bands
$r^{\prime}$ and $i^{\prime}$.

\begin{figure*}
\centering 
\includegraphics[width=18cm]{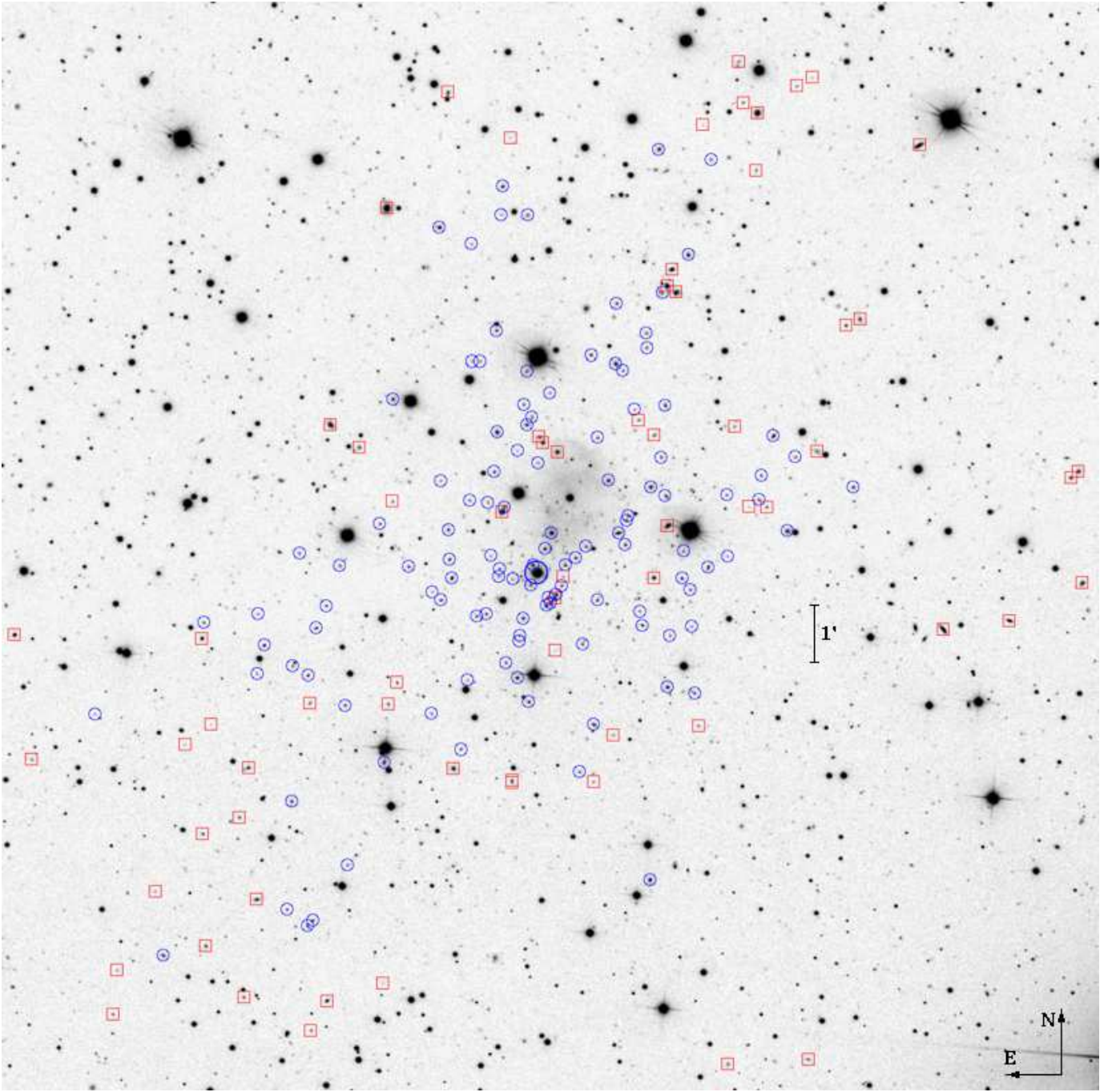}
\caption{Wide field SDSS image (filter $r^{\prime}$) of the cluster CL1821 
  showing the positions of the galaxies of our spectroscopic
  catalogue. Blue circles and red squares indicate cluster members and
  non-member galaxies, respectively (see
  Table~\ref{catalogCL1821}). The biggest circle in the centre
  highlights the central quasar H1821+643 (ID~94 and BCG; see the text).}
\label{figottico}
\end{figure*}

Table~\ref{catalogCL1821} lists the velocity catalogue (see also
Fig.~\ref{figottico}): identification number of each galaxy, ID
(Column~1); equatorial coordinates right ascension and declination,
$\alpha$ and $\delta$ (J2000, Column~2); (dereddened) SDSS
$r^{\prime}$ magnitude (Column~3); heliocentric radial velocities,
$V=cz$ (Column~4) with errors; $\Delta V$ (Column~5).


\addtocounter{table}{0}
\begin{table}
        \caption[]{Radial velocities of 187 galaxies in the field of
          CL1821. \dagg highlights the velocities taken from the literature
          (see the text). IDs in italics refer to non-member
          galaxies. Galaxy ID~94 is the BCG.}
         \label{catalogCL1821}
              $$ 
           \begin{array}{r c c r r}
            \hline
            \noalign{\smallskip}
            \hline
            \noalign{\smallskip}

\mathrm{ID} & \alpha,\delta\,(\mathrm{J}2000) & r^{\prime}\, &V\,& \Delta V\\
 &     (18^{\rm h},+64^{\rm o})             & &\mathrm{\,(\,km}&\mathrm{s^{-1}\,)}\\
            \hline
            \noalign{\smallskip}  

\textit{1}   &     20\ 31.59, 20\ 24.1 & 18.68 &    \dag75389 & 72\\  
\textit{2}   &     20\ 32.09, 22\ 19.0 & 18.84 &    \dag83609 &114\\  
\textit{3}   &     20\ 33.26, 22\ 12.2 & 19.37 &    \dag99630 & 40\\  
\textit{4}   &     20\ 43.06, 19\ 45.9 & 18.33 &    \dag26771 & 67\\  
\textit{5}   &     20\ 53.44, 19\ 36.9 & 17.80 &    \dag33442 &100\\  
\textit{6}   &     20\ 56.88, 27\ 55.5 & 17.72 &    \dag21495 & 27\\  
\textit{7}   &     21\ 06.39, 24\ 56.5 & 18.89 &    73611 &110\\  
 8           &     21\ 07.52, 22\ 03.7 & 19.46 &    89599 & 90\\  
\textit{9}   &     21\ 08.58, 24\ 50.1 & 19.70 &    73745 & 90\\  
\textit{10}  &     21\ 13.25, 22\ 40.8 & 19.24 &    56522 & 95\\  
\textit{11}  &     21\ 13.80, 29\ 05.2 & 21.54 &   148779 &100\\  
\textit{12}  &     21\ 14.79, 12\ 15.3 & 19.95 &    \dag55474 & 35\\  
\textit{13}  &     21\ 16.22, 28\ 56.4 & 20.46 &   117818 &117\\  
14           &     21\ 16.59, 22\ 34.8 & 20.13 &    88890 &106\\  
15           &     21\ 17.90, 21\ 18.8 & 18.83 &    \dag89248 &100\\  
16           &     21\ 20.04, 22\ 56.7 & 18.89 &    \dag87587 & 53\\  
\textit{17}  &     21\ 21.15, 21\ 43.6 & 20.24 &   \dag110084 &100\\  
18           &     21\ 21.99, 22\ 15.9 & 20.11 &    88209 &154\\  
19           &     21\ 22.38, 21\ 51.1 & 19.82 &    86861 & 90\\  
\textit{20}  &     21\ 22.47, 28\ 29.1 & 17.33 &    36706 & 55\\  
\textit{21}  &     21\ 22.67, 27\ 29.9 & 20.14 &    79645 &180\\  
\textit{22}  &     21\ 23.90, 21\ 43.8 & 21.57 &   \dag164346 &100\\  
\textit{23}  &     21\ 24.73, 28\ 39.5 & 20.43 &   146570 & 88\\  
\textit{24}  &     21\ 25.39, 29\ 21.7 & 21.02 &   117742 &112\\  
\textit{25}  &     21\ 26.19, 23\ 06.4 & 20.29 &   130358 &108\\  
26           &     21\ 27.31, 20\ 52.7 & 20.94 &    89229 &103\\  
\textit{27}  &     21\ 27.41, 12\ 10.5 & 19.76 &    \dag8250 & 12\\  
28           &     21\ 27.44, 21\ 55.9 & 20.36 &    \dag89878 &100\\  
29           &     21\ 29.79, 27\ 41.0 & 20.01 &    89915 &132\\  
30           &     21\ 30.35, 20\ 41.7 & 19.48 &    \dag87210 &100\\  
\textit{31}  &     21\ 31.09, 28\ 16.6 & 21.75 &    85853 &154\\  
\textit{32}  &     21\ 31.93, 17\ 58.5 & 20.44 &   146525 &143\\  
33           &     21\ 32.55, 18\ 32.0 & 20.02 &    88987 & 99\\  
34           &     21\ 32.98, 19\ 40.7 & 20.77 &    88993 &123\\  
35           &     21\ 33.17, 20\ 18.6 & 20.69 &    89495 & 77\\  
36           &     21\ 33.43, 26\ 03.0 & 18.88 &    89826 & 84\\  
37           &     21\ 34.24, 20\ 58.2 & 20.56 &    \dag88918 &100\\  
38           &     21\ 34.48, 20\ 30.2 & 19.84 &    \dag87072 & 82\\  
\textit{39}  &     21\ 35.39, 25\ 24.6 & 18.19 &    \dag56661 &100\\  
\textit{40}  &     21\ 36.00, 25\ 48.2 & 18.27 &    56836 & 42\\  
41           &     21\ 36.37, 19\ 31.4 & 20.43 &    88725 &106\\  
\textit{42}  &     21\ 36.71, 21\ 24.6 & 18.21 &    \dag51223 & 74\\  
\textit{43}  &     21\ 36.73, 25\ 31.1 & 18.20 &    67816 & 66\\  
                                                                  
            \noalign{\smallskip}			          
            \hline					    
            \noalign{\smallskip}			    
            \hline					    
         \end{array}					 
     $$ 						 
         \end{table}					 
\addtocounter{table}{-1}				 
\begin{table}					 
          \caption[ ]{Continued.}
     $$ 
           \begin{array}{r c c r r}
            \hline
            \noalign{\smallskip}
            \hline
            \noalign{\smallskip}
\mathrm{ID} & \alpha,\delta\,(\mathrm{J}2000) & r^{\prime}\, &V\,& \Delta V\\
 &     (18^{\rm h},+64^{\rm o})     & &\mathrm{\,(\,km}&\mathrm{s^{-1}\,)}\\

            \hline
            \noalign{\smallskip}

44           &     21\ 36.76, 18\ 38.5 &  19.06 &  89491 & 66\\ 
45           &     21\ 37.07, 21\ 55.8 &  20.44 &  85899 &108\\ 
46           &     21\ 37.07, 23\ 28.4 &  19.33 &  88842 & 88\\ 
47           &     21\ 37.51, 25\ 23.9 &  20.38 &  91479 &101\\ 
48           &     21\ 37.72, 22\ 35.0 &  19.93 &  85741 & 75\\ 
49           &     21\ 38.00, 27\ 51.8 &  18.88 &  89440 &103\\ 
\textit{50}  &     21\ 38.79, 22\ 57.8 &  19.69 &  67852 & 79\\ 
\textit{51}  &     21\ 38.91, 20\ 30.9 &  19.10 &  \dag67903 & 32\\ 
52           &     21\ 39.37, 22\ 04.3 &  18.82 &  \dag86580 & 92\\ 
53           &     21\ 39.57, 15\ 20.3 &  18.69 &  \dag91524 & 50\\ 
54           &     21\ 39.99, 24\ 27.7 &  19.88 &  90396 & 80\\ 
55           &     21\ 40.02, 24\ 42.5 &  19.60 &  88670 &145\\ 
56           &     21\ 40.71, 19\ 42.2 &  19.56 &  89368 & 88\\ 
57           &     21\ 41.13, 19\ 56.6 &  20.67 &  86653 &110\\ 
\textit{58}  &     21\ 41.25, 23\ 13.4 &  20.17 &  49290 &110\\ 
59           &     21\ 41.87, 23\ 24.0 &  20.65 &  88200 &167\\ 
60           &     21\ 42.90, 21\ 35.0 &  20.13 &  86930 &147\\ 
61           &     21\ 43.18, 21\ 30.0 &  20.35 &  85307 &136\\ 
62           &     21\ 43.37, 21\ 05.0 &  19.53 &  88883 &103\\ 
63           &     21\ 43.71, 24\ 04.0 &  20.20 &  88092 & 84\\ 
64           &     21\ 44.46, 21\ 17.2 &  18.76 &  89154 & 59\\ 
65           &     21\ 44.69, 25\ 13.3 &  19.78 &  88496 &130\\ 
66           &     21\ 44.86, 24\ 11.1 &  18.57 &  87976 & 79\\ 
\textit{67}  &     21\ 45.33, 17\ 49.2 &  20.72 & 153152 &121\\ 
68           &     21\ 46.02, 22\ 11.1 &  19.01 &  89467 & 75\\ 
69           &     21\ 47.71, 22\ 55.1 &  20.00 &  87825 &114\\ 
70           &     21\ 47.78, 20\ 07.9 &  19.89 &  \dag87539 &100\\ 
71           &     21\ 48.33, 18\ 00.0 &  18.93 &  89988 & 88\\ 
\textit{72}  &     21\ 48.41, 17\ 01.2 &  20.77 &  56519 &143\\ 
73           &     21\ 48.75, 24\ 20.3 &  19.55 &  86423 & 88\\ 
74           &     21\ 49.54, 21\ 03.1 &  20.40 &  89747 & 88\\ 
75           &     21\ 50.10, 19\ 22.7 &  20.16 &  87940 &238\\ 
76           &     21\ 50.58, 17\ 11.4 &  20.04 &  89298 & 90\\ 
77           &     21\ 51.16, 20\ 51.5 &  20.04 &  \dag88079 &100\\ 
78           &     21\ 52.82, 20\ 43.9 &  19.71 &  \dag90387 &300\\ 
\textit{79}  &     21\ 53.14, 20\ 32.0 &  21.16 & \dag119917 &300\\ 
80           &     21\ 53.36, 20\ 23.0 &  20.22 &  \dag87108 & 83\\ 
\textit{81}  &     21\ 54.06, 22\ 40.5 &  18.46 &  53682 & 97\\ 
82           &     21\ 54.25, 20\ 13.7 &  19.70 &  \dag89788 &300\\ 
\textit{83}  &     21\ 54.34, 19\ 16.6 &  22.23 & 262551 &394\\ 
\textit{84}  &     21\ 54.44, 20\ 09.6 &  19.11 &  \dag67633 &150\\ 
85           &     21\ 54.96, 21\ 17.2 &  18.23 &  \dag91137 &300\\ 
86           &     21\ 55.09, 20\ 04.8 &  19.36 &  \dag87839 &100\\ 

            \noalign{\smallskip}			    
            \hline					    
            \noalign{\smallskip}			    
            \hline					    
         \end{array}
     $$ 
         \end{table}
\addtocounter{table}{-1}
\begin{table}
          \caption[ ]{Continued.}
     $$ 
           \begin{array}{r c c r r}
            \hline
            \noalign{\smallskip}
            \hline
            \noalign{\smallskip}

\mathrm{ID} & \alpha,\delta\,(\mathrm{J}2000) & r^{\prime}\, &V\,& \Delta V\\
 &    (18^{\rm h},+64^{\rm o})    & &\mathrm{\,(\,km}&\mathrm{s^{-1}\,)}\\

            \hline
            \noalign{\smallskip}
   
87           &     21\ 55.20, 23\ 41.2 &  20.38 &   88807 &158\\ 
88           &     21\ 55.56, 20\ 10.4 &  20.19 &   89103 &114\\ 
89           &     21\ 55.80, 20\ 02.7 &  19.69 &   \dag90819 & 55\\ 
90           &     21\ 56.00, 21\ 00.9 &  19.22 &   \dag87419 &150\\ 
\textit{91}  &     21\ 56.35, 22\ 50.2 &  18.94 &   \dag73254 & 64\\ 
\textit{92}  &     21\ 56.92, 22\ 55.9 &  19.82 &   73731 & 44\\ 
93           &     21\ 57.10, 22\ 29.4 &  20.67 &   91369 &125\\ 
94           &     21\ 57.24, 20\ 36.2 &  13.60 &   89054 & 48\\ 
95           &     21\ 57.86, 20\ 44.2 &  19.79 &   \dag90327 &100\\ 
96           &     21\ 58.06, 23\ 16.2 &  20.34 &   90691 &119\\ 
97           &     21\ 58.26, 20\ 23.3 &  19.98 &   88271 &130\\ 
98           &     21\ 58.53, 18\ 23.3 &  19.61 &   \dag87288 & 89\\ 
99           &     21\ 58.73, 26\ 44.2 &  19.62 &   \dag88595 & 84\\ 
100          &     21\ 58.83, 24\ 03.2 &  20.96 &   90203 &132\\ 
101          &     21\ 58.86, 23\ 08.4 &  19.36 &   91283 & 84\\ 
102          &     21\ 59.35, 23\ 29.2 &  20.20 &   88972 &136\\ 
103          &     21\ 59.41, 19\ 49.3 &  19.48 &   \dag88289 &300\\ 
104          &     21\ 59.99, 19\ 31.2 &  20.84 &   86866 &117\\ 
105          &     22\ 00.02, 19\ 26.2 &  21.44 &   89579 &114\\ 
106          &     22\ 00.35, 22\ 41.6 &  20.75 &   91662 &139\\ 
107          &     22\ 00.36, 18\ 48.1 &  19.04 &   86260 & 59\\ 
108          &     22\ 01.09, 20\ 29.6 &  21.21 &   \dag88439 &100\\ 
\textit{109} &     22\ 01.13, 17\ 00.6 &  21.16 &   80169 &121\\ 
\textit{110} &     22\ 01.21, 17\ 02.5 &  19.63 &   79486 &136\\ 
\textit{111} &     22\ 01.48, 28\ 03.7 &  21.55 &  178877 &108\\ 
112          &     22\ 02.14, 19\ 03.4 &  20.27 &   89787 & 95\\ 
113          &     22\ 02.40, 21\ 43.7 &  19.75 &   89200 &103\\ 
114          &     22\ 02.69, 27\ 13.7 &  19.20 &   89219 &121\\ 
\textit{115} &     22\ 02.71, 21\ 38.9 &  18.10 &   36368 & 86\\ 
116          &     22\ 02.85, 26\ 44.5 &  20.71 &   89022 &150\\ 
117          &     22\ 03.13, 20\ 40.3 &  20.37 &   88322 &114\\ 
118          &     22\ 03.26, 20\ 32.6 &  20.36 &   91075 & 66\\ 
119          &     22\ 03.53, 23\ 00.9 &  18.57 &   \dag89236 & 80\\ 
120          &     22\ 03.63, 24\ 45.3 &  19.64 &   87826 & 81\\ 
121          &     22\ 03.95, 22\ 20.0 &  19.93 &   85679 & 53\\ 
122          &     22\ 04.55, 20\ 53.5 &  20.77 &   \dag85261 &100\\ 
123          &     22\ 04.96, 21\ 48.6 &  20.56 &   89694 &103\\ 
124          &     22\ 05.23, 19\ 53.9 &  19.99 &   88874 &101\\ 
125          &     22\ 06.20, 24\ 14.1 &  20.60 &   89703 &163\\ 
126          &     22\ 06.72, 19\ 51.4 &  19.94 &   \dag90477 &100\\ 
127          &     22\ 07.44, 24\ 14.0 &  20.82 &   86153 &154\\ 
128          &     22\ 07.57, 26\ 14.3 &  20.75 &   90618 & 95\\ 
129          &     22\ 07.86, 21\ 50.6 &  20.33 &   90209 & 88\\ 
                                                     
            \noalign{\smallskip}			    
            \hline					    
            \noalign{\smallskip}			    
            \hline					    
         \end{array}
     $$ 
         \end{table}
\addtocounter{table}{-1}
\begin{table}
          \caption[ ]{Continued.}
     $$ 
           \begin{array}{r c c r r}
            \hline
            \noalign{\smallskip}
            \hline
            \noalign{\smallskip}

\mathrm{ID} & \alpha,\delta\,(\mathrm{J}2000) & r^{\prime}\, &V\,& \Delta V\\
 &    (18^{\rm h},+64^{\rm o})    & &\mathrm{\,(\,km}&\mathrm{s^{-1}\,)}\\

            \hline
            \noalign{\smallskip}
   
130          &     22\ 08.15, 18\ 45.5 &  21.13 &    87170 &141\\ 
131          &     22\ 09.19, 17\ 34.7 &  20.08 &    86584 & 66\\ 
\textit{132} &     22\ 10.36, 17\ 14.8 &  18.49 &    \dag79952 & 80\\ 
133          &     22\ 10.63, 20\ 30.7 &  19.01 &    \dag87449 &100\\ 
134          &     22\ 10.98, 20\ 49.8 &  19.65 &    86298 & 59\\ 
135          &     22\ 11.19, 21\ 20.4 &  19.55 &    90179 &101\\ 
\textit{136} &     22\ 11.33, 28\ 50.3 &  19.81 &    \dag99138 &144\\ 
137          &     22\ 12.33, 20\ 07.9 &  19.68 &    88022 & 68\\ 
138          &     22\ 12.41, 22\ 10.9 &  20.63 &    88765 &169\\ 
139          &     22\ 12.62, 26\ 31.4 &  18.67 &    \dag89422 & 66\\ 
140          &     22\ 13.67, 20\ 16.5 &  20.89 &    \dag89008 &100\\ 
141          &     22\ 13.79, 18\ 11.3 &  20.51 &    87989 &134\\ 
142          &     22\ 17.42, 20\ 42.3 &  19.76 &    \dag89938 &100\\ 
\textit{143} &     22\ 19.21, 18\ 43.1 &  20.18 &    \dag73653 & 86\\ 
144          &     22\ 19.93, 23\ 34.7 &  19.27 &    87897 & 51\\ 
\textit{145} &     22\ 19.97, 21\ 49.4 &  20.50 &    \dag53513 &100\\ 
\textit{146} &     22\ 20.59, 18\ 20.8 &  20.19 &   155238 &134\\ 
\textit{147} &     22\ 20.96, 26\ 51.0 &  16.64 &    \dag57596 & 35\\ 
148          &     22\ 21.24, 17\ 21.1 &  19.43 &    87092 &150\\ 
\textit{149} &     22\ 21.39, 13\ 33.7 &  21.67 &   126279 & 90\\ 
150          &     22\ 21.94, 21\ 26.4 &  20.11 &    89185 & 62\\ 
\textit{151} &     22\ 25.21, 22\ 44.9 &  19.52 &    \dag56862 & 40\\ 
152          &     22\ 26.95, 15\ 35.4 &  20.36 &    89798 &136\\ 
153          &     22\ 27.37, 18\ 19.6 &  19.94 &    88821 &121\\ 
154          &     22\ 28.22, 20\ 43.2 &  19.96 &    \dag88529 &100\\ 
\textit{155} &     22\ 29.75, 23\ 08.2 &  18.05 &    \dag57446 & 53\\ 
\textit{156} &     22\ 30.13, 13\ 15.6 &  18.89 &    \dag83903 & 55\\ 
157          &     22\ 30.35, 20\ 01.9 &  20.03 &    90509 & 92\\ 
158          &     22\ 31.94, 19\ 39.0 &  19.74 &    \dag90238 &100\\ 
159          &     22\ 32.36, 14\ 38.2 &  20.00 &    88410 &125\\ 
\textit{160} &     22\ 32.61, 12\ 45.2 &  20.39 &    84424 &233\\ 
\textit{161} &     22\ 32.88, 18\ 22.0 &  20.09 &    73594 &141\\ 
162          &     22\ 33.20, 18\ 50.4 &  19.68 &    \dag88382 & 54\\ 
163          &     22\ 33.21, 14\ 32.9 &  19.71 &    90483 &114\\ 
164          &     22\ 34.61, 20\ 55.9 &  20.07 &    \dag90238 &100\\ 
165          &     22\ 35.63, 18\ 59.8 &  20.38 &    89943 &123\\ 
166          &     22\ 35.70, 16\ 40.7 &  19.47 &    90309 &123\\ 
167          &     22\ 36.39, 14\ 49.6 &  20.11 &    87349 &128\\ 
168          &     22\ 40.09, 19\ 21.2 &  19.45 &    87967 & 88\\ 
169          &     22\ 41.06, 19\ 53.2 &  20.59 &    90338 & 75\\ 
\textit{170} &     22\ 41.16, 14\ 59.8 &  19.02 &    57626 & 75\\ 
171          &     22\ 41.21, 18\ 52.0 &  20.53 &    89142 &108\\ 
\textit{172} &     22\ 42.50, 17\ 15.1 &  19.55 &    72691 & 84\\ 
                                                    
            \noalign{\smallskip}			    
            \hline					    
            \noalign{\smallskip}			    
            \hline					    
         \end{array}
     $$ 
         \end{table}
\addtocounter{table}{-1}
\begin{table}
          \caption[ ]{Continued.}
     $$ 
           \begin{array}{r c c r r}
            \hline
            \noalign{\smallskip}
            \hline
            \noalign{\smallskip}

\mathrm{ID} & \alpha,\delta\,(\mathrm{J}2000) & r^{\prime}\, &V\,& \Delta V\\
 &    (18^{\rm h},+64^{\rm o})    & &\mathrm{\,(\,km}&\mathrm{s^{-1}\,)}\\

            \hline
            \noalign{\smallskip}
   
\textit{173} &     22\ 43.12, 13\ 19.0 & 20.09  &    84216 &106\\ 
\textit{174} &     22\ 43.89, 16\ 24.1 & 19.95  &   130560 &123\\ 
\textit{175} &     22\ 48.41, 17\ 59.5 & 21.96  &   224389 &103\\ 
\textit{176} &     22\ 49.09, 14\ 11.6 & 19.60  &    80114 & 84\\ 
177          &     22\ 49.60, 19\ 44.4 & 20.12  &    90348 & 95\\ 
\textit{178} &     22\ 49.63, 16\ 06.9 & 19.97  &    82326 &123\\ 
\textit{179} &     22\ 49.85, 19\ 28.1 & 18.93  &    49129 & 88\\ 
\textit{180} &     22\ 52.49, 17\ 38.7 & 21.49  &   167915 &157\\ 
181          &     22\ 55.75, 14\ 01.7 & 19.31  &    89898 &128\\ 
\textit{182} &     22\ 57.06, 15\ 07.6 & 20.92  &   110810 &114\\ 
\textit{183} &     23\ 02.99, 13\ 46.2 & 20.93  &    78316 &163\\ 
\textit{184} &     23\ 03.60, 13\ 01.2 & 20.40  &    65097 & 86\\ 
185          &     23\ 06.64, 18\ 09.9 & 21.01  &    88598 &161\\ 
\textit{186} &     23\ 16.49, 17\ 22.6 & 20.11  &   147851 &156\\ 
\textit{187} &     23\ 19.29, 19\ 31.0 & 18.30  &   \dag172860 &300\\ 
                                                    
            \noalign{\smallskip}			    
            \hline					    
            \noalign{\smallskip}			    
            \hline					    
         \end{array}
     $$ 
\end{table}

The cluster galaxy population is dominated by the quasar H1821+643
(ID~94 and hereafter BCG; see its optical spectrum in
Fig.~\ref{figimage}). Although classified as a radio-quiet quasar, the
BCG is a Fanaroff--Riley type I source (Blundell \&
Rawlings \citeyear{blu01}), and is evident in high-resolution radio
images of CL1821 taken by B14 (see their fig.~1) and at X-ray
wavelengths. ID~85 is the second brightest cluster member in our
catalogue and is also an FR-I (head--tail) radio galaxy (Lacy et
al. \citeyear{lac92}; Blundell \& Rawlings \citeyear{blu01}). B14
report a bright and elongated radio source $\sim$2.9 arcmin SW of BCG
(their source ``D''), at only $\sim$3 arcsec from the member galaxy
ID~44, which is the probable optical counterpart. Among other galaxies
with radio and/or X-ray emission in the field of CL1821, our galaxy
ID~83 is a background active galactic nucleus with
z$\sim$0.88. Fainter radio sources (from B14) appear associated to our
galaxies ID~68 (a cluster member) and 84 and 115 (foreground
non-member galaxies).

\section{ANALYSIS OF THE SAMPLE}
\label{optanalysis}

\subsection{Members selection and global properties}
\label{memb}

\begin{figure}
\centering
\resizebox{\hsize}{!}{\includegraphics{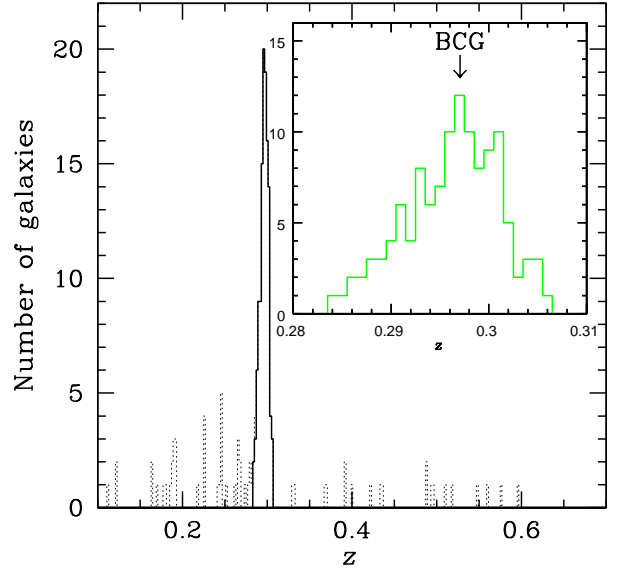}}
\caption
{Redshift galaxy distribution. The heavy line histogram refers to the
  131 galaxies assigned to CL1821 according to the 1D-DEDICA
  reconstruction method. The inset figure shows the member-galaxy
  distribution (after final selection with the ``shifting gapper''
  method) with the indication of the BCG redshift.}
\label{fighisto}
\end{figure}

\begin{figure}
\centering
\resizebox{\hsize}{!}{\includegraphics{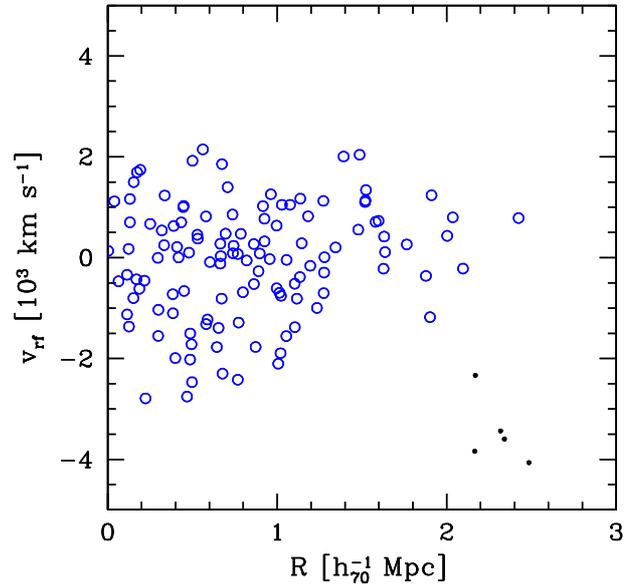}}
\caption
{Rest-frame velocity versus projected cluster-centric distance for the 120
  member galaxies (blue circles). Black dots are galaxies selected by
  1D-DEDICA but rejected by the ``shifting gapper'' test.}
\label{figvd}
\end{figure}

The selection of cluster members was performed using two statistical
tools: the 1D adaptive-kernel method (hereafter 1D-DEDICA,
Pisani \citeyear{pis93}) and the ``shifting gapper'' method (Fadda et
al. \citeyear{fad96}). First, we ran 1D-DEDICA on the 187 galaxies of
our spectroscopic catalogue and detected CL1821 as a prominent velocity
peak at $z\sim 0.296$ populated by 131 provisional cluster members
(see Fig~\ref{fighisto}). In the second step, we combined the spatial
and velocity information for the provisional members by running the
``shifting gapper'' test. In particular, we chose a gap of 1000 \ks
(see also Girardi et al. \citeyear{gir96}) and adopted for the cluster
centre the location of the BCG. With this tool we rejected another
eleven galaxies leading to a final sample of 120 cluster members,
whose projected phase space is shown in Fig.~\ref{figvd}.

The velocity distribution of the 120 cluster members was analysed with
the biweight routines by Beers et al. (\citeyear{bee90}), which
provide robust estimates of location and scale. Our measurement of the
mean cluster redshift is $\left<z\right>=0.2965\pm0.0003$ (i.e.,
$\left<V\right>=88\,878\pm104$ \kss). Instead, the global LOS velocity
dispersion is $\sigma_{\rm V}=1137_{-78}^{+72}$ \kss. Adopting the
$\sigma_{\rm V}$-$M_{200}$ relation of Munari et
al. (\citeyear{mun13}; derived from cosmological $N$-body and
hydrodynamic simulations in the $\Lambda$-cold dark matter framework),
we compute a total cluster mass of $M_{200} = (1.4 \pm 0.2)\times
10^{15}$\m within $R_{200}=2.1$ Mpc.

\subsection{Cluster substructures}
\label{sub}

We searched for substructures by using 1D (velocity information), 2D
(spatial distribution), and 3D (combined velocity and spatial
information) statistical tests. 

The 1D analysis consists of a detailed study of the properties of the
velocity distribution of member galaxies (Fig.~\ref{fighisto}). In
particular, we do not find significant evidence of deviations from
Gaussianity (a possible hint of complex dynamics in the cluster)
according to several parameters (kurtosis, skewness, and tail
index). Instead, there is marginal evidence of asymmetry according to
the W-Test ($\sim 95$\% c.l.) and the asymmetry index ($\sim 90-95$\%
c.l.; see Bird \& Beers \citeyear{bir93} for details).

We also ran the 1D-Kaye Mixture Model test (1D-KMM; Ashman et
al. \citeyear{ash94}; see also for recent applications Girardi et
al. \citeyear{gir08} and Balestra et al. \citeyear{bal16}) to search for
bimodal partitions of the velocity distribution fitting the data
better than a single Gaussian. In the homoscedastic case
(i.e. assuming equal velocity dispersions), the most likely solution
(at the 95\% c.l.) identifies two peaks at 87124 \ks and 89428 \ks
with $\sigma_{\rm V}\sim 800 \pm 100$ \kss populated by 29 and 91
galaxies, respectively. According to the 2D Kolmogorov-Smirnov test, it
is not possible to separate the two peaks in the spatial
domain. Instead, in the physically more realistic (heteroscedastic,
i.e. assuming different velocity dispersions) case, the 1D-KMM test
does not provide any significant solution.

%
%

\begin{figure}
\resizebox{\hsize}{!}{\includegraphics{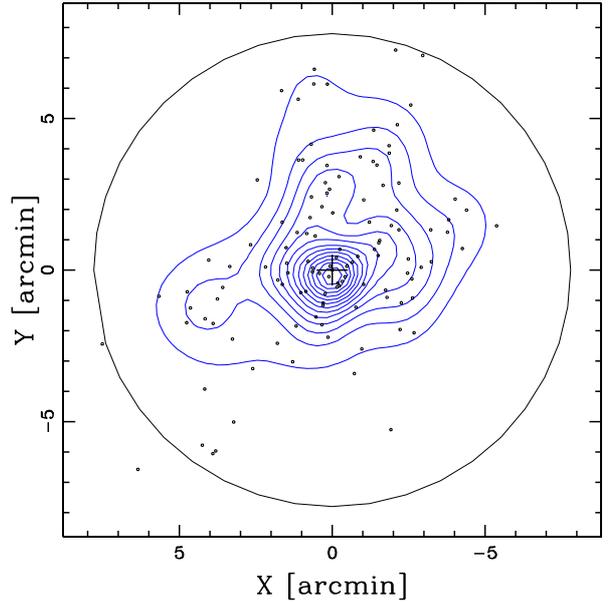}}
\caption{Spatial distribution on the sky of
  the 120 spectroscopic cluster members and relative isodensity
  contour map obtained with the 2D-DEDICA method. The black cross
  indicates the position of BCG. The plot is centred on the cluster
  centre and circle contains the cluster within a radius equal to
  7.8 arcmin$\sim 2.1$ \h ($\sim$ $R_{200}$).}
\label{figk2g}
\end{figure}

About the analysis of the 2D spatial distribution of the spectroscopic
member galaxies, we employed the 2D adaptive-kernel method of Pisani
et al. (\citeyear{pis96}, hereafter 2D-DEDICA). The results are shown
in Fig.~\ref{figk2g}. This test detects CL1821 as a very prominent
galaxy peak with two low-density secondary clumps: the densest one is
located $\sim$2 arcmin N of the cluster centre (density only $\sim$35\%
the main peak) and the second one lies $\sim$4 arcmin ESE of the cluster
centre (see also Table~\ref{tabdedica2dz}). However, the existence of
these two low-density subclumps appears very questionable, also taking
into account that our spectroscopic data do not cover the entire
cluster field and are affected by magnitude incompleteness due to
unavoidable constraints in the design of the MOS masks.

\begin{figure}
\resizebox{\hsize}{!}{\includegraphics{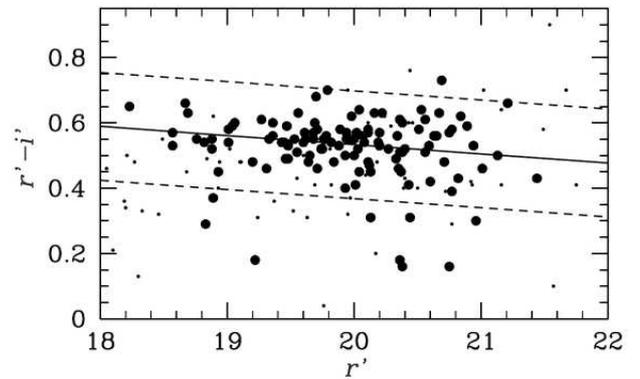}}
\caption{$r^{\prime}-i^{\prime}$ versus $r^{\prime}$ diagram
  for galaxies with available spectroscopy. Big circles and dots
  indicate member and non-member galaxies, respectively. The solid line
  gives the CMR determined on member galaxies; the dashed lines are
  drawn at color index $\pm$0.16 mag from the CMR (see the text).}
\label{figcm}
\end{figure}

We can overcome our incompleteness problems by using the SDSS
photometry of the cluster field at our disposal. In particular, we
consider all the galaxies within 3 Mpc from the cluster centre and
select likely members on the basis of the ($r^{\prime}-i^{\prime}$
versus $r^{\prime}$) colour-magnitude relation (hereafter CMR), which
allows to identify the locus of the dominant cluster population
(i.e. early-type galaxies, Dressler \citeyear{dre80}). We determine
the CMR by applying the 2$\sigma$-clipping fitting procedure to the
cluster members and obtain
$r^{\prime}$--$i^{\prime}$=1.093-0.028$\times r^{\prime}$ (see
Fig.~\ref{figcm}). Then, within the photometric catalogue we consider
as likely ``red'' cluster members the galaxies with colour index
within 0.16 mag of the CMR (i.e. the 1$\sigma$-error associated to the
fitted intercept).

\begin{figure}
\includegraphics[width=8cm]{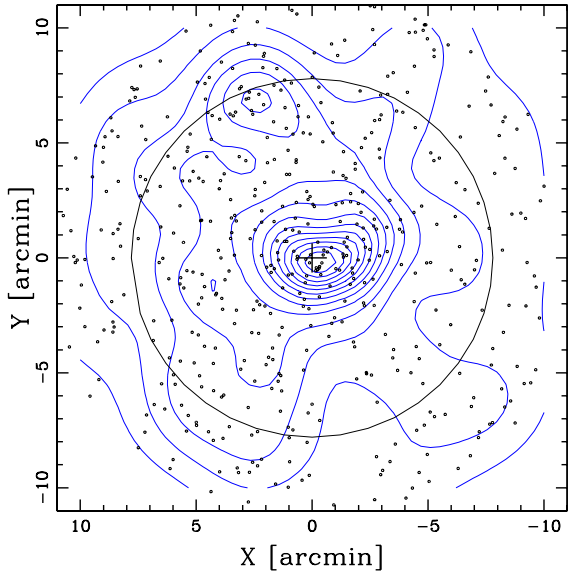}
\caption{Spatial distribution on the sky and relative isodensity contour map
  of the photometric likely cluster members with $r^{\prime}\le 21$
  obtained with the 2D-DEDICA test. The plot is centred on the
  cluster centre and circle contains the cluster within a radius equal
  to 7.8 arcmin $\sim 2.1$ \h ($\sim$ $R_{200}$).}
\label{figk2}
\end{figure}

Figure~\ref{figk2} shows the contour map for the 508 likely cluster
members having $r^{\prime}\le 21$ according to 2D-DEDICA. There is no
trace now of the two subclumps found with the previous analysis, thus
suggesting that they are artefacts due to the incompleteness of the
spectroscopic sample. Instead, there is evidence of a more distant
subgroup lying $\sim$2 Mpc ($\sim R_{200}$) NNE of the cluster centre
(see Table~\ref{tabdedica2dphot}).

\begin{figure}
\includegraphics[width=8cm]{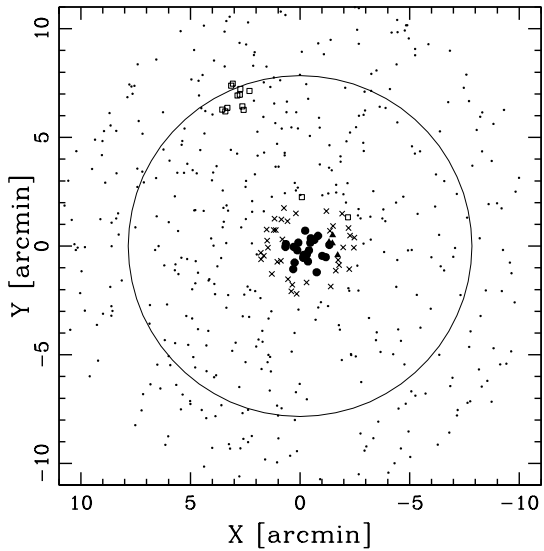}
\caption {Galaxies belonging to structures as detected by the VTP method.
The algorithm is run on the sample of likely members with
$r^{\prime}\le 21$ (see the text). Open squares, crosses, triangles
and solid circles indicate galaxies in structures at the 90, 95, 98,
99\% c. ls., respectively. The plot is centred on the cluster centre
and circle contains the cluster within a radius equal to 7.8 arcmin
$\sim 2.1$ \h ($\sim$ $R_{200}$).}
\label{figvtp}
\end{figure}

The existence of this external structure is also confirmed by the
Voronoi Tessellation and Percolation (VTP) technique (e.g. Ramella et
al. \citeyear{ram01}; Barrena et al. \citeyear{bar05}), a
non-parametric test which is sensitive to galaxy structures regardless
of their shapes. Considering again the 508 likely members with
$r^{\prime}\le 21$, we ran VTP over this set four times adopting four
detection thresholds. The results are shown in Fig.~\ref{figvtp},
where galaxies identified as belonging to structures at 90, 95,
98, and 99\% c.ls. are drawn as open squares, crosses, triangles,
and solid circles, respectively. Once again VTP detects the main body
of CL1821 without hints of substructures and the new external group at
NNE, even though the last one is at the lowest c.l.

\begin{table}
        \caption[]{The 2D structure of CL1821. Results from the spectroscopic sample according to the 2D-DEDICA method. For each subclump the table lists the number of assigned member galaxies $N_{\rm S}$, equatorial coordinates of the density peak, the relative density with respect to the highest peak $\rho_{\rm S}$, and the $\chi^2$ value of the peak.}
         \label{tabdedica2dz}
            $$
         \begin{array}{l r c c r }
            \hline
            \noalign{\smallskip}
            \hline
            \noalign{\smallskip}
\mathrm{Subclump} & N_{\rm S} & \alpha({\rm J}2000),\,\delta({\rm J}2000)&\rho_{
\rm S}&\chi^2_{\rm S}\\
& & \mathrm{h:m:s,\degree:\arcmm:\arcs}&&\\
         \hline
         \noalign{\smallskip}
\mathrm{C}      & 72&18\ 21\ 57.1,+64\ 20\ 25&1.00&33.3\\
\mathrm{N}      & 30&18\ 21\ 59.2,+64\ 23\ 01&0.35&12.8\\
\mathrm{ESE}    & 13&18\ 22\ 34.5,+64\ 19\ 24&0.17& 6.2\\
              \noalign{\smallskip}
              \noalign{\smallskip}
            \hline
            \noalign{\smallskip}
            \hline
         \end{array}
$$
\end{table}

\begin{table}
        \caption[]{The 2D structure of CL1821. Results from the sample
          of photometric likely members according to the 2D-DEDICA
          method. For each subclump the table lists the number of
          assigned member galaxies $N_{\rm S}$, equatorial coordinates of the
          density peak, the relative density with respect to the
          highest peak $\rho_{\rm S}$, and the $\chi^2$ value of the
          peak.}
         \label{tabdedica2dphot}
            $$
         \begin{array}{l r c c r }
            \hline
            \noalign{\smallskip}
            \hline
            \noalign{\smallskip}
\mathrm{Subclump} & N_{\rm S} & \alpha({\rm J}2000),\,\delta({\rm J}2000)&\rho_{
\rm S}&\chi^2_{\rm S}\\
& & \mathrm{h:m:s,\degree:\arcmm:\arcs}\\
         \hline
         \noalign{\smallskip}
\mathrm{C}      &330&18\ 21\ 54.3,+64\ 20\ 24&1.00&50.3\\
\mathrm{NNE}    & 67&18\ 22\ 20.0,+64\ 27\ 28&0.45&16.8\\

              \noalign{\smallskip}
              \noalign{\smallskip}
            \hline
            \noalign{\smallskip}
            \hline
         \end{array}
$$
\end{table}

%
%

As for the 3D analysis, we employed different tools to search for a
correlation between velocity and position information, which would be
a clear sign of real substructures in the cluster.

First, we find no evidence of a significant velocity gradient (see
e.g. den Hartog \& Katgert \citeyear{den96} and Girardi et
al. \citeyear{gir96} for the details of the method) in the sample of
the 120 spectroscopic cluster members.

Then, over the same sample we apply the classical $\Delta$-test
(Dressler \& Schectman \citeyear{dre88}, hereafter DS test), which is
considered as the most sensitive 3D test (e.g. Pinkney et
al. \citeyear{pin96}). For each spectroscopic member galaxy, the
DS test computes the deviation $\delta$ defined as $\delta^2
= \delta_{\rm V}^2 + \delta_{\rm S}^2$, where $\delta_{\rm
V}^2=[(N_{\rm nn}+1)/\sigma_{\rm V}^2]\times (\overline {\rm V_{loc}}
- \overline {\rm V})^2$, $\delta_{\rm S}^2= [(N_{\rm
nn}+1)/\sigma_{\rm V}^2]\times (\sigma_{\rm V,loc} - \sigma_{\rm
V})^2$ and the subscript ``loc'' denotes the local quantities computed
over the set containing the galaxy and its $N_{\rm{nn}}=10$
neighbours. The parameter $\Delta$ is then defined as the sum of the
$\delta$ of the individual galaxies and provides the cumulative
deviation of the local kinematical parameters (mean velocity and
velocity dispersion) from the global cluster parameters. The
significance of $\Delta$, i.e. of substructure, is finally checked by
running Monte Carlo simulations.

No significant substructure is found in CL1821 with the DS test and
its modified versions, which consider separately the local mean
velocity (parameter $\delta_{\rm V}^2$) and velocity dispersion
(parameter $\delta_{\rm S}^2$) as kinematical indicators (see Girardi
et al. \citeyear{gir10}).

Finally, we resorted to the hierarchical tree (Htree) algorithm
developed by Serna and Gerbal (\citeyear{ser96}; see also Girardi et
al. \citeyear{gir11} and Durret et al. \citeyear{dur15} for recent
applications). We apply it to the catalogue of 120 member galaxies
with known radial velocities and $r^{\prime}$-band magnitudes. The
method executes a hierarchical clustering analysis and extracts galaxy
subgroups by computing the relative binding energies of member
galaxies. It has proved to be effective to detect substructures both
in nearby clusters (e.g.  Coma -- Adami et al. \citeyear{ada05}) and
in medium-redshift clusters (e.g. Abell 1995 -- Boschin et
al. \citeyear{bos12a}). It works with a constant value for the
mass-to-light ratio of galaxies. Here we adopt a value of
$M/L_{r^{\prime}}$=150 \ml (a value comparable to that of clusters,
see Serna \& Gerbal \citeyear{ser96}), but the gross results are quite
robust against the choice of the value of $M/L_{r^{\prime}}$. We also
assume a magnitude $r^{\prime}\sim +17$ for the host galaxy (the BCG)
of the central quasar H1821+643 (from measurements performed by
Hutchings and Neff \citeyear{hut91} and Floyd et al. \citeyear{flo04})
and choose $n=10$ as the minimum number of galaxies in substructures
(e.g. Durret et al. \citeyear{dur15}). The Htree method does not find
significant substructure in CL1821.

\section{DISCUSSION AND CONCLUSIONS}
\label{disc}

The high values of the velocity dispersion $\sigma_{\rm
V}=1137_{-78}^{+72}$ \ks and X-ray temperature $kT_{\rm
X}=(9.0\pm0.5)$ keV (excluding the central cool core; Russell et
al. \citeyear{rus10}) of CL1821 are typical of hot, massive clusters
(e.g. Girardi \& Mezzetti \citeyear{gir01}). Moreover, our estimate of
$\sigma_{\rm V}$ and $T_{\rm X}$ are consistent each other with the
assumption of energy density equipartition between ICM and
galaxies. In fact, we obtain $\beta_{\rm spec} = \sigma_{\rm
V}^2/(kT_{\rm X}/\mu m_{\rm p}) = 0.87^{+0.13}_{-0.12}$, to be
compared with $\beta_{\rm spec}=1$\footnote{$\mu=0.58$ is the mean
molecular weight and $m_{\rm p}$ the proton mass.}. This differs from
what is commonly found in merging clusters hosting GRHs, where values
of $\beta_{\rm spec}$ significantly greater than unity are often found
(e.g. Abell 2744 -- Boschin et al. \citeyear{bos06}; Abell 2254 --
Girardi et al. \citeyear{gir11}; Abell 1758N -- Boschin et
al. \citeyear{bos12b}; Abell 1351 -- Barrena et al. \citeyear{bar14}).

However, the most convincing evidence that CL1821 is not interested in
a major merger comes from the absence of significant substructures
according to the analysis of our optical data.

In particular, the analysis of the 2D galaxy distribution performed
with the 2D-DEDICA test detects CL1821 as a very prominent and well-
isolated galaxy peak. The only significant subclump is the one located
at a projected distance of $\sim$2 Mpc NNE of the cluster centre and
detected on the sample of ``likely'' members both with 2D-DEDICA (see
Fig.~\ref{figk2} and Table~\ref{tabdedica2dphot}) and with the VTP
method (Fig.~\ref{figvtp}). Unfortunately, we have very few redshifts
in the region of the subgroup, and its full characterization is not
possible. However, its projected galaxy density makes of it a minor
system at $R_{200}$ from the cluster centre, which does not seem able
to perturb significantly the potential well of CL1821.

These findings in the spatial domain agree, as expected, with the lack
of major substructures observed in {\it Chandra} X-ray data (B14;
Russell et al. \citeyear{rus10}). However, our main goal was to
explore the existence of ongoing mergers along the LOS. The 1D
velocity distribution of member galaxies would provide a decisive
contribution in this sense. Our results do not show strong evidence of
departure from Gaussianity and asymmetry, which would be important
signs of a dynamically perturbed cluster. The 1D-KMM test proposes a
marginally significant partition of the velocity distribution in two
peaks, but only in the physically (homoscedastic) less realistic
case. In the heteroscedastic case, on the contrary, 1D-KMM does not
find any significant solution for a bimodal (or more complex)
partition of the velocity distribution.

Finally, combining the velocity and position information of member
galaxies we do not find evidence of a significant velocity
gradient. This suggests that eventual substructures would be mainly
LOS aligned, but the results of the DS test and the Htree method agree
with the negative findings of the 1D and 2D analysis.

In conclusion, our new optical data provide a picture of CL1821
consistent with being a massive dynamically relaxed cluster dominated
by a BCG located in the centre of the cluster both in the velocity
(see Fig.~\ref{fighisto}) and in the spatial domains. According to our
results, the presence of a cool core in CL1821 is not surprising, even
though a cool core is not necessarily disrupted if a cluster is
affected by a merger with enough angular momentum (see
e.g. simulations by Poole et al. \citeyear{poo08}; Hahn et
al. \citeyear{hah17}).

Nevertheless, the absence of major substructures within 1.5-2 Mpc from
the cluster centre is in conflict with the existence of the GRH
discovered in CL1821.

Indeed, a major merger is not always needed to produce diffuse radio
emission in a cluster. In fact, minor mergers are also predicted to
form haloes with ultrasteep radio spectrum and low-radio power with
respect to major mergers (e.g., Cassano et al. \citeyear{cas06};
Brunetti et al. \citeyear{bru08}), but leaving the cluster core
intact. This could be the case of PSZ1G139.61+24 mentioned in
Sect.~\ref{intro} (Savini et al. \citeyear{sav18}), a cool-core
cluster with an associated mini-halo and a low-size ($\sim$0.5 Mpc)
and underluminous radio halo with a spectral index\footnote{We define
the radio spectrum as $S(\nu)\varpropto \nu^{-\alpha}$.}
$\alpha \gtrsim 1.7$. However, the GRH hosted by CL1821 and the halo
found in PSZ1G139.61+24 can be hardly compared with each other,
especially considering that the GRH of CL1821 resembles more a
``classical'' radio halo, with a flatter spectrum ($\alpha \sim
1.0-1.1$) and a size of $\gtrsim 1$ Mpc (see B14).

Instead, PSZ1G139.61+24 shares some similarities with Abell 2142,
another cluster which is not a major merger but presents a
two-component diffuse radio emission (Venturi et
al. \citeyear{ven17}). Its central brightest component has a spectral
index $\alpha$ similar to those of typical radio haloes and Venturi et
al. (\citeyear{ven17}) propose that it could be powered by the gas
sloshing in the centre of the cluster (similar to the scenario
suggested for the formation of radio mini-haloes in cool-core
clusters; see e.g., Mazzotta \& Giacintucci \citeyear{maz08}; ZuHone
et al. \citeyear{zuh13}) with a central AGN activity which would
provide seed electrons then reaccelerated in the ICM. If such a
mechanism could play any role in explaining also the GRH of CL1821 is
questionable. In fact, in terms of overall size and spectral steepness
the GRH of CL1821 could partially resemble the diffuse emission in
Abell 2142. Moreover, the central BCG (and perhaps also the second
brightest member ID~85; see Sect.~\ref{prom}) could act as the AGNs
located in the centre of Abell 2142. However, the differences between
the two clusters are evident: CL1821 is a cool-core cluster and,
according to our analysis, does not show obvious signs of recent
mergers, while Abell 2142 is not a cool-core cluster and exhibits
multiple cold fronts which are signs of past minor merger activity.

In conclusion, the existence of the GRH hosted by CL1821 remains an
open problem, which brings into question our current understanding of
cool-core clusters and diffuse radio emission in clusters.

\section*{Acknowledgments}

We thank the anonymous referee for his/her useful comments and
suggestions.

MG acknowledges the support from the grant MIUR PRIN 2015 ``Cosmology
and Fundamental Physics: illuminating the Dark Universe with Euclid''.

This publication is based on observations made on the island of La
Palma with the Italian Telescopio Nazionale Galileo (TNG), which is
operated by the Fundaci\'on Galileo Galilei -- INAF (Istituto
Nazionale di Astrofisica) and is located in the Spanish Observatorio
of the Roque de Los Muchachos of the Instituto de Astrof\'isica de
Canarias.

This research has made use of the galaxy catalogue of the Sloan
Digital Sky Survey (SDSS). The SDSS web site is http://www.sdss.org/,
where the list of the funding organizations and collaborating
institutions can be found.





\begin{thebibliography}{99}

\bibitem[\protect\citeauthoryear{Adami et al.}{2005}]{ada05} Adami C., Biviano A., Durret F, Mazure A., 2005, A\&A, 443, 17

\bibitem[\protect\citeauthoryear{Aravena et al.}{2011}]{ara11} Aravena M., Wagg J., Papadopoulos P. P., Feain I. J., 2011, ApJ, 737, 64

\bibitem[\protect\citeauthoryear{Ashman et al.}{1994}]{ash94} Ashman K. M., Bird C. M., Zepf S. E., 1994, AJ, 108, 2348

\bibitem[\protect\citeauthoryear{Balestra et al.}{2016}]{bal16} Balestra I., Mercurio A., Sartoris B., et al., 2016, ApJS, 224, 33

\bibitem[\protect\citeauthoryear{Barrena et al.}{2014}]{bar14} Barrena, R., Girardi M., Boschin W., De Grandi S., Rossetti M., 2014, MNRAS, 442, 2216

\bibitem[\protect\citeauthoryear{Barrena et al.}{2005}]{bar05} Barrena R., Ramella M., Boschin W., et al., 2005, A\&A, 444, 685

\bibitem[\protect\citeauthoryear{Beers et al.}{1990}]{bee90} Beers T. C., Flynn K., Gebhardt K., 1990, AJ, 100, 32

\bibitem[\protect\citeauthoryear{Bird \& Beers}{1993}]{bir93} Bird C. M., Beers, T. C., 1993, AJ, 105, 1596

\bibitem[\protect\citeauthoryear{Blundell \& Rawlings}{2001}]{blu01} Blundell K. M., Rawlings S., 2001, ApJ, 562, L5

\bibitem[\protect\citeauthoryear{Bonafede et al.}{2014}]{bon14} Bonafede, A., Intema, H. T., Br\"uggen, M., et al., 2014, MNRAS, 444, L44 (B14)

\bibitem[\protect\citeauthoryear{Boschin et al.}{2013}]{bos13} Boschin W., Girardi M., Barrena R., 2013, \mnras, 434, 772

\bibitem[\protect\citeauthoryear{Boschin et al.}{2004}]{bos04} Boschin W., Girardi M., Barrena R., et al., 2004, A\&A, 416, 839

\bibitem[\protect\citeauthoryear{Boschin et al.}{2012a}]{bos12a} Boschin W., Girardi M., Barrena R., et al., 2012a, A\&A, 547, A44

\bibitem[\protect\citeauthoryear{Boschin et al.}{2012b}]{bos12b} Boschin W., Girardi M., Barrena R., Nonino M., 2012b, A\&A, 540, A43

\bibitem[\protect\citeauthoryear{Boschin et al.}{2006}]{bos06} Boschin W., Girardi M., Spolaor M., Barrena R., et al., 2006, A\&A, 449, 461


\bibitem[\protect\citeauthoryear{Brunetti \& Jones}{2015}]{bru15} Brunetti G., Jones T. W., 2015, {\it Magnetic Fields in Diffuse Media}, Astrophysics and Space Science Library, Vol. 407, Springer-Verlag Berlin Heidelberg, p. 557

\bibitem[\protect\citeauthoryear{Brunetti et al.}{2008}]{bru08} Brunetti G., Giacintucci S., Cassano R., et al., 2008, Nature, 455, 944

\bibitem[\protect\citeauthoryear{Cassano et al.}{2006}]{cas06} Cassano R., Brunetti G., Setti G., 2006, MNRAS, 369, 1577

\bibitem[\protect\citeauthoryear{den Hartog \& Katgert}{1996}]{den96} den Hartog R., Katgert P., 1996, MNRAS, 279, 349

\bibitem[\protect\citeauthoryear{Dressler}{1980}]{dre80} Dressler A., 1980, ApJ, 236, 351 

\bibitem[\protect\citeauthoryear{Dressler \& Schectman}{1988}]{dre88} Dressler A., Shectman S. A., 1988, AJ, 95, 985

\bibitem[\protect\citeauthoryear{Durret et al.}{2015}]{dur15} Durret F., Wakamatsu K., Nagayama T., Adami C., Biviano A., 2015, A\&A, 583, A124

\bibitem[\protect\citeauthoryear{Fadda et al.}{1996}]{fad96} Fadda D., Girardi M., Giuricin G., Mardirossian F., Mezzetti M., 1996, ApJ, 473, 670

\bibitem[\protect\citeauthoryear{Feretti et al.}{2012}]{fer12} Feretti L., Giovannini G., Govoni F., Murgia M., 2012, A\&ARv, 20, 54

\bibitem[\protect\citeauthoryear{Floyd et al.}{2004}]{flo04} Floyd D. J. E., Kukula M. J., Dunlop J. S., et al., 2004, MNRAS, 355, 196

\bibitem[\protect\citeauthoryear{Girardi et al.}{2011}]{gir11} Girardi M., Bardelli S., Barrena R., et al., 2011, A\&A, 536, A89

\bibitem[\protect\citeauthoryear{Girardi et al.}{2008}]{gir08} Girardi M., Barrena R., Boschin W., Ellingson E., 2008, A\&A, 491, 379

\bibitem[\protect\citeauthoryear{Girardi et al.}{2010}]{gir10} Girardi M., Boschin W., Barrena R., 2010, A\&A, 517, A65

\bibitem[\protect\citeauthoryear{Girardi et al.}{1996}]{gir96} Girardi M., Fadda D., Giuricin G., et al., 1996, ApJ, 457, 61

\bibitem[\protect\citeauthoryear{Girardi \& Mezzetti}{2001}]{gir01} Girardi M., Mezzetti M., 2001, ApJ, 548, 79

\bibitem[\protect\citeauthoryear{Hahn et al.}{2017}]{hah17} Hahn O., Martizzi D., Wu H.-Y., et al., 2017, MNRAS, 470, 166

\bibitem[\protect\citeauthoryear{Hutchings \& Neff}{1991}]{hut91} Hutchings J. B., Neff S. G., 1991, AJ, 101, 2001

\bibitem[\protect\citeauthoryear{Kale \& Parekh}{2016}]{kal16} Kale R., Parekh V., 2016, MNRAS, 459, 2940

\bibitem[\protect\citeauthoryear{Lacy et al.}{1992}]{lac92} Lacy M., Rawlings S., Hill G. J., 1992, MNRAS, 258, 828

\bibitem[\protect\citeauthoryear{Mazzotta \& Giacintucci}{2008}]{maz08} Mazzotta P., Giacintucci S., 2008, ApJ, 675, L9

\bibitem[\protect\citeauthoryear{Munari et al.}{2013}]{mun13} Munari E., Biviano A., Borgani S., Murante G., Fabjan D., 2013, MNRAS, 430, 2638

\bibitem[\protect\citeauthoryear{Parekh et al.}{2015}]{par15} Parekh V., van der Heyden K., Ferrari C., Angus G., Holwerda B., 2015, A\&A, 575, A127

\bibitem[\protect\citeauthoryear{Pinkney et al.}{1996}]{pin96} Pinkney J., Roettiger K., Burns J. O., Bird C. M., 1996, \apjs, 104, 1

\bibitem[\protect\citeauthoryear{Pisani}{1993}]{pis93} Pisani A., 1993, MNRAS, 265, 706

\bibitem[\protect\citeauthoryear{Pisani}{1996}]{pis96} Pisani A., 1996, MNRAS, 278, 697

\bibitem[\protect\citeauthoryear{Poole et al.}{2008}]{poo08} Poole G. B., Babul A., McCarthy I. G., Sanderson A. J. R., Fardal M. A., 2008, MNRAS, 391, 1163 

\bibitem[\protect\citeauthoryear{Ramella et al.}{2001}]{ram01} Ramella M., Boschin W., Fadda D., Nonino M., 2001, A\&A, 368, 776

\bibitem[\protect\citeauthoryear{Reynolds et al.}{2014}]{rey14}	Reynolds C. S., Lohfink A. M., Babul A., et al., 2014, ApJ, 792, L41

\bibitem[\protect\citeauthoryear{Roettiger et al.}{1997}]{roe97} Roettiger K., Loken C., Burns J. O., 1997, ApJS, 109, 307

\bibitem[\protect\citeauthoryear{Russell et al.}{2010}]{rus10} Russell H. R., Fabian A. C., Sanders J. S., et al., 2010, MNRAS, 402, 1561

\bibitem[\protect\citeauthoryear{Savini}{2018}]{sav18} Savini F., Bonafede A., Br\"uggen M., et al., 2018, MNRAS, 478, 2234

\bibitem[\protect\citeauthoryear{Schneider}{1992}]{sch92} Schneider D. P., Bahcall J. N., Gunn J. E., Dressler A., 1992, AJ, 103, 1047

\bibitem[\protect\citeauthoryear{Serna \& Gerbal}{1996}]{ser96} Serna A., Gerbal D., 1996, A\&A, 309, 65

\bibitem[\protect\citeauthoryear{Sommer et al.}{2017}]{som17} Sommer M. W., Basu K., Intema H., et al., 2017, MNRAS, 466, 966 

\bibitem[\protect\citeauthoryear{Tonry \& Davis}{1979}]{ton79} Tonry J., Davis M., 1979, ApJ, 84, 1511

\bibitem[\protect\citeauthoryear{Venturi}{2017}]{ven17} Venturi T., Rossetti M., Brunetti G., et al., 2017, A\&A, 603, A125

\bibitem[\protect\citeauthoryear{Walker et al.}{2014}]{wal14} Walker S. A., Fabian A. C., Russell H. R., Sanders J. S., 2014, MNRAS, 442, 2809

\bibitem[\protect\citeauthoryear{Wold et al.}{2002}]{wol02} Wold M., Lacy M., Dahle H., Lilje P. B., Ridgway S. E., 2002, MNRAS, 335, 1017

\bibitem[\protect\citeauthoryear{ZuHone et al.}{2013}]{zuh13} ZuHone J., Markevitch M., Brunetti G., et al., 2013, ApJ, 762, 78

\end{thebibliography}


\bsp	
\label{lastpage}
\end{document}